\documentclass[
  aps,
  prb,
  showpacs,
  superscriptaddress,
  floatfix,
  twocolumn,
  secnumarabic,
  amssymb,
  amsmath]{revtex4-1}

\usepackage{graphicx}
\usepackage{bm}
\usepackage{color}
\usepackage[colorlinks=true,citecolor=blue,linkcolor=blue,pdfstartview=FitH,pdfauthor=Gepraegs]{hyperref}
\begin{document}

\title{Investigation of induced Pt magnetic polarization in Pt/Y$_3$Fe$_5$O$_{12}$ bilayers}

\author{Stephan Gepr\"{a}gs}
 \email{Stephan.Gepraegs@wmi.badw.de}
 \affiliation{Walther-Mei{\ss}ner-Institut, Bayerische Akademie der
              Wissenschaften, 85748 Garching, Germany}

\author{Sibylle Meyer}
 \affiliation{Walther-Mei{\ss}ner-Institut, Bayerische Akademie der
              Wissenschaften, 85748 Garching, Germany}

\author{Stephan Altmannshofer}
 \affiliation{Walther-Mei{\ss}ner-Institut, Bayerische Akademie der
              Wissenschaften, 85748 Garching, Germany}

\author{Matthias Opel}
 \affiliation{Walther-Mei{\ss}ner-Institut, Bayerische Akademie der
              Wissenschaften, 85748 Garching, Germany}

\author{Fabrice Wilhelm}
 \affiliation{European Synchrotron Radiation Facility (ESRF), 6 Rue Jules Horowitz, BP 220, 38043 Grenoble Cedex 9, France}

\author{Andrei Rogalev}
 \affiliation{European Synchrotron Radiation Facility (ESRF), 6 Rue Jules Horowitz, BP 220, 38043 Grenoble Cedex 9, France}

\author{Rudolf Gross}
 \affiliation{Walther-Mei{\ss}ner-Institut, Bayerische Akademie der
              Wissenschaften, 85748 Garching, Germany}
 \affiliation{Physik-Department, Technische Universit\"{a}t M\"{u}nchen, 85748 Garching, Germany}

\author{Sebastian T. B.~Goennenwein}
 \affiliation{Walther-Mei{\ss}ner-Institut, Bayerische Akademie der
              Wissenschaften, 85748 Garching, Germany}

\date{\today}%

\begin{abstract}

Using X-ray magnetic circular dichroism (XMCD) measurements, we explore the possible existence of induced magnetic moments in thin Pt films deposited onto the ferrimagnetic insulator yttrium iron garnet (Y$_3$Fe$_5$O$_{12}$). Such a magnetic proximity effect is well established for Pt/ferromagnetic metal heterostructures. Indeed, we observe a clear XMCD signal at the Pt $L_3$ edge in Pt/Fe bilayers, while no such signal can be discerned in XMCD traces of Pt/Y$_3$Fe$_5$O$_{12}$ bilayers. Integrating the XMCD signals allows to estimate an upper limit for the induced Pt magnetic polarization in Pt/Y$_3$Fe$_5$O$_{12}$ bilayers.
\end{abstract}


\maketitle

Pure spin currents are a fascinating manifestation of spin physics in the solid state.\cite{Hirsch:PRL:1999,Stevens:PRL:2003,Kato:Science:2004,Saitoh:APL:2006,Mosendz:PRL:2010,Uchida:NatMater:2010} Experimentally, the generation or detection of spin currents is often based on the interconversion of spin and charge currents, taking advantage of the spin Hall or the inverse spin Hall effect, respectively. \cite{Hirsch:PRL:1999,Kato:Science:2004,Saitoh:APL:2006,Uchida:NatMater:2010} This makes normal metal/ferromagnetic metal (NM/FMM) or normal metal/ferromagnetic insulator (NM/FMI) heterostructures very attractive.\footnote{For the sake of simplicity, we use the term 'ferromagnetic' for both ferro- and ferrimagnetic materials throughout this paper. A 'ferromagnetic insulator' thus is an electrically insulating material which exhibits long-range ferro- or ferri-magnetic ordering.} In so-called spin pumping\cite{Tserkovnyak:PRL:2002,Saitoh:APL:2006,Mosendz:PRL:2010,Czeschka:PRL:2011} or spin Seebeck experiments,\cite{Uchida:Nature:2008,Uchida:NatMater:2010,Jaworski:NatMater:2010,Weiler:PRL:2012,Jaworski:Nature:2012} the magnetization in the ferromagnetic constituent is driven out of thermal equilibrium, and the ensuing spin current into the normal metal (NM) layer is detected via the corresponding inverse spin Hall electrical current in these heterostructures. Hereby, the paramagnetic NM layer is commonly considered as 'non-magnetic' in the sense that its spin polarization is negligibly small, such that magneto-resistive or magneto-thermo-galvanic effects in the NM layer can be safely ignored.\cite{Uchida:NatMater:2010,Uchida:APL:2010} This assumption appears reasonable for NM/FMM bilayers, in which the unavoidable magneto-thermo-galvanic response of the FMM layer dominates. In contrast, for the case of NM/FMI structures, the absence of an induced spin polarization in the NM layer and thus the complete absence of magneto-thermo-galvanic effects in the FMI such as the anomalous Nernst effect are considered as an advantage and even exploited for the interpretation.\cite{Uchida:NatMater:2010,Uchida:APL:2010} Very recently, magnetic proximity effects in NM/FMI structures were inferred from electrical and thermal magnetotransport measurements.\cite{Huang:PRL:2012} Thus, a careful investigation of a possible finite induced magnetic polarization in the NM layer in NM/FMI heterostructures is essential for the correct interpretation of spin current related phenomena. 
In NM/FMM heterostructures and alloys using $3d$ and $5d$ elements, the presence of an induced spin polarization in the NM layer in proximity to the interface has been observed by X-ray magnetic circular dichroism (XMCD) experiments.\cite{Wilhelm:PRL:2000,Wilhelm:PRL:2001,Wilhelm:PRB:2004,Wende:RepProgPhys:2004,Parra:PRB:1980} To our knowledge, no such investigations have been performed in NM/FMI heterostructures. 

In this letter, we report on XMCD measurements in NM/FMI bilayers, composed of the 'non-magnetic' metal Pt and the ferrimagnetic insulator yttrium iron garnet (Y$_{3}$Fe$_{5}$O$_{12}$, YIG). For comparison, we also investigated a Pt/Fe (NM/FMM) bilayer as a reference sample. While a clear XMCD signal at the Pt $L_{3}$ edge in Pt/Fe unambiguously demonstrates the presence of spin-polarized electrons in the Pt layer, no such signal could be resolved in within the sensitivity of the experiment. However, integrating the XMCD signal in the sample with the thinnest Pt layer studied, we estimate an upper limit for the induced total magnetic moment, which is at least 30 times smaller than in the Pt/Fe reference sample. A similar evaluation in Pt/YIG samples with thicker Pt layers yields no finite integrated XMCD signal.

A series of Pt/YIG bilayers was fabricated by laser-MBE and subsequent electron-beam evaporation in ultra-high vacuum on (111)-oriented, single crystalline Y$_{3}$Al$_{5}$O$_{12}$ (YAG) substrates. First, the YIG thin film was epitaxially grown via pulsed laser deposition (PLD) from a stoichiometric, polycrystalline target using a KrF excimer laser ($\lambda=$248\,nm) with a laser fluence of 2\,J/cm$^{2}$ and a repetition rate of 10\,Hz. The deposition was carried out in an oxygen atmosphere at a pressure of 25\,$\mu$bar and a substrate temperature of $500^\circ$C. Second, the polycrystalline Pt layer was deposited \textit{in-situ} without breaking the vacuum on top of the YIG thin film using electron-beam evaporation. We here investigate three samples with similar thickness of the YIG thin film ($\sim\!62$\,nm) and different thicknesses (3\,nm, 7\,nm, 10\,nm) of the Pt layer. For comparison, we fabricated a polycrystalline Pt(10\,nm)/Fe(10\,nm) reference sample solely by electron-beam evaporation in ultra-high vacuum.

All bilayer samples were characterized with respect to their structural and magnetic properties. X-ray diffraction measurements reveal no secondary phases in the thin films. Furthermore, the rocking curves around the YIG(444) reflection display a full width at half maximum of about 0.1°, indicating a moderate mosaic spread of the YIG thin films. This can be mainly attributed to the lattice mismatch between the YIG thin film and the YAG substrate of about 3\%. However, the surface roughness of the YIG thin films was found to be lower than 1\,nm (rms value). The magnetic properties were investigated using superconducting quantum interference device (SQUID) magnetometry. For plain YIG thin films deposited on YAG substrates, we obtained a saturation magnetization of $M=110$\,kA/m close to the bulk value of $M_{\mathrm{s}}^{\mathrm{YIG}}=143$\,kA/m\cite{Coey:book} at room temperature. In the Pt/YIG bilayers, $M_{\mathrm{s}}$ is further reduced, which might be explained as a result of interdiffusion of Fe and Al at the interface between the YIG thin film and the YAG substrate during the deposition of Pt.\cite{Popova:EPJB:2003}

%
\begin{figure}[tb]
  \includegraphics[]{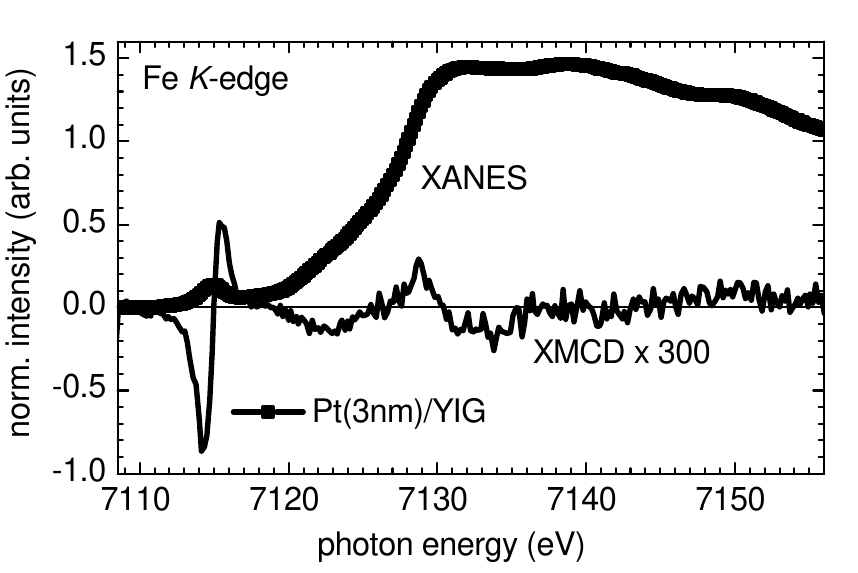}
    \caption{Normalized XANES (symbols) and corresponding normalized XMCD signal (line) from a Pt(3\,nm)/YIG bilayer at the Fe $K$ edge. The XMCD spectrum is multiplied by a factor of 300 for clarity.}
    \label{fig:Fig1}
\end{figure}

To further investigate the magnetic properties of the Pt/YIG bilayers in particular with respect to induced magnetic moments in the Pt layer, we present a comprehensive study taking advantage of the element-specific XMCD technique in the following. The measurements were performed at the European Synchrotron Radiation Facility (ESRF) at the beam line ID12 using the total fluorescence yield (TFY) detection mode. X-ray absorption near edge spectra (XANES) were recorded at the Fe $K$ (7112\,eV) and the Pt $L_3$ edge (11564\,eV) with right and left circularly polarized light as well as positive and negative magnetic fields applied. An electro-magnet, which allows to flip the magnetic field direction at each point of the incoming photon energy, was used to apply magnetic fields of $\pm 60$\,mT parallel to the direction of the incoming X-ray beam. The samples were placed in the center of the magnet under a small angle of around $15^\circ$ between their surface and the incoming light. Several spectra were recorded to improve the signal-to-noise ratio and normalized to an edge jump of unity. The normalized and averaged spectra were then used to calculate the XMCD signal as the direct difference between consecutive XANES recorded either with right and left circularly polarized light or while applying positive and negative magnetic fields. We note that the XMCD measurements presented in the following therefore give access only to the projection of the magneticzation onto the external magnetic field (viz.~X-ray beam) direction. A possible magnetic moment perpendicular to the X-ray beam direction is then hardly detectable.

Figure~\ref{fig:Fig1} shows the normalized XANES and the corresponding XMCD spectrum at the Fe $K$ edge from the Pt(3\,nm)/YIG bilayer. The XANES exhibits two notable features: the pre-edge peak located at $\sim\!7115$\,eV and the main edge at $\sim\!7130$\,eV.\cite{Rogalev:JMMM:2009} The pre-edge structure, which is attributed to Fe$^{3+}$ ions occupying the tetrahedrally coordinated sites,\cite{Kawamura:JPhysIV:1997,Maruyama:ElecSpec:2004} dominates the XMCD signal (cf.~line in Fig.~\ref{fig:Fig1}). Comparing the signals with corresponding spectra from YIG single crystals\cite{Rogalev:JMMM:2009} reveals no qualitative differences. This fact further demonstrates the high quality of our YIG thin films, in particular with respect to the Fe coordination.

\begin{figure}[tbh]
  \includegraphics[]{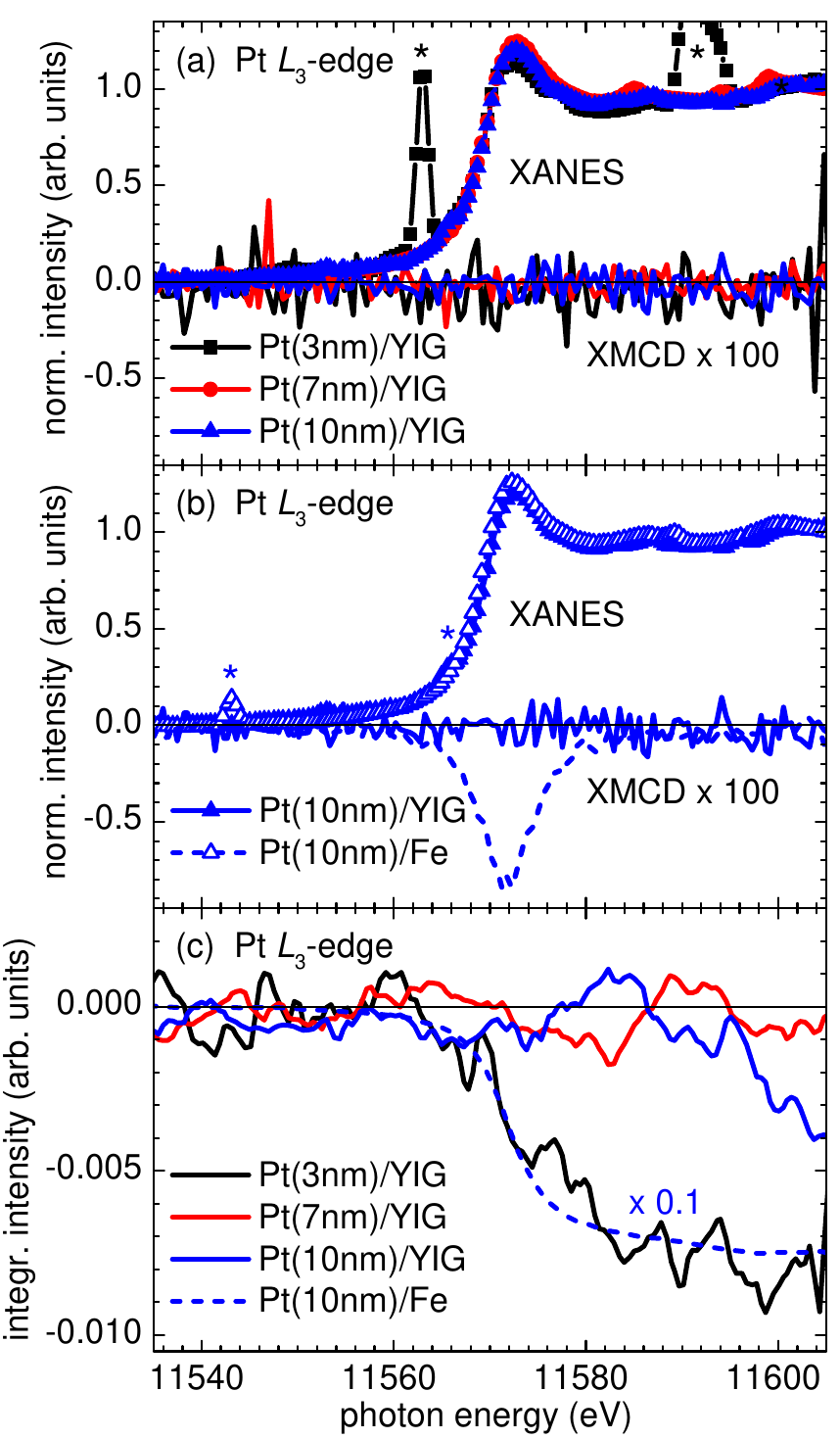}
    \caption{(color online) (a) Normalized Pt $L_{3}$ edge XANES (symbols) and XMCD (lines) spectra from Pt/YIG bilayers with different Pt thicknesses: 3\,nm (black), 7\,nm (red), and 10\,nm (blue). Diffraction peaks are marked by asterisks. (b) Corresponding XANES and XMCD signal from a Pt(10\,nm)/Fe bilayer (open symbols and dashed line). The normalized XANES (full symbols) and XMCD spectrum (solid line) from the Pt(10\,nm)/YIG bilayer is shown for comparison. For better illustration, all XMCD spectra are multiplied by a factor of 100. (c) Corresponding integrated XMCD intensity. Before the integration, a small background slope was subtracted from the spectra. The integrated intensity of the Pt(10\,nm)/Fe bilayer was multiplied by a factor of 0.1 (dashed line).}
    \label{fig:Fig2}
\end{figure}

To probe a possible induced magnetic polarization in the Pt layer of Pt/YIG (NM/FMI) and Pt/Fe (NM/FMM) samples, XANES are recorded at the Pt $L_3$ edge, which is known to show a larger XMCD signal than the $L_2$ edge.\cite{Wilhelm:PRL:2000} Figure~\ref{fig:Fig2}(a) displays the normalized XANES (symbols) for three Pt/YIG bilayers with different thicknesses of the Pt layer. Some spurious peaks are caused by elastic diffraction from the YAG substrate and marked by asterisks. The fine structure of the XANES is often used as a 'fingerprint' for possible oxidation of the Pt layer.\cite{Mansour:JPhysChem:1984,Friebel:PhysChemChemPhys:2011,Miller:PRL:2011} However, from the recorded XANES, no clear evidence for the formation of platinum oxide, which becomes apparent in a stronger white-line intensity and differences in the intensity profile of the post-edge region, could be found in our Pt thin films deposited on top of YIG. Thus, outward diffusion of oxygen from the YIG thin film into the Pt layer as well as oxidation effects at the Pt surface can be discarded.

After averaging up to 34 single XANES per Pt/YIG sample with different helicity of the X-rays and different magnetic field directions, the XMCD signal for each bilayer was calculated [cf.~lines in Fig.~\ref{fig:Fig2}(a)]. As evident from Fig.~\ref{fig:Fig2}(a), the XMCD spectra do not show indications for a finite XMCD signal at the Pt $L_{3}$ edge, not even, apparently, in the sample with the thinnest Pt layer exhibiting the largest interface-to-volume ratio.

To cross-check our experimental approach and to confirm earlier XMCD results from Pt/Ni multilayers\cite{Wilhelm:PRL:2000} or PtFe nanoparticles,\cite{Antoniak:PRL:2006} XANES and XMCD spectra were recorded at the Pt $L_{3}$ edge from the Pt(10\,nm)/Fe(10\,nm) reference sample. Figure~\ref{fig:Fig2}(b) reveals almost no difference between the XANES from Pt(10\,nm)/Fe (cf.~open symbols) and from Pt(10\,nm)/YIG (cf.~full symbols). This further demonstrates, that no alteration of the Pt layer due to oxidation takes place in the Pt/YIG bilayers with respect to the Pt/Fe sample. 
However, in contrast to the Pt/YIG bilayer [cf.~solid line in Fig.~\ref{fig:Fig2}(b)], a clear XMCD signal is detected in the Pt/Fe bilayer [cf.~dashed line in Fig.~\ref{fig:Fig2}(b)]. 
This result unambiguously demonstrates the presence of induced magnetic moments in the Pt layer deposited on Fe, which is in agreement with earlier XMCD results.\cite{Wilhelm:PRL:2000,Wende:RepProgPhys:2004} 
%
%
The data published in the literature also shows that the moments induced in Pt are basically located close to the Pt/FMM interface, while the induced polarization rapidly decays exponentially with increasing Pt thickness.\cite{Poulopoulos:JAP:2001} However, as obvious from Fig.~\ref{fig:Fig2}(b), a clear XMCD signal could be resolved even in Pt/Fe bilayers with a large thickness of the Pt layer of 10\,nm. To extract the total moment $\mu_{t}$ of Pt via the standard magneto-optical sum rules, the XMCD intensity is integrated [cf.~dashed line in Fig.~\ref{fig:Fig2}(c)] and a ratio of the orbital magnetic moment $\mu_{l}$ and spin magnetic moment $\mu_{s}$ of $\mu_{l}/\mu_{s}=0.131$\cite{Antoniak:PRL:2006} is assumed.\footnote{The ratio $\mu_{l}/\mu_{s}$ for induced moments in Pt in direct contact with a ferromagnetic late $3d$ transition metal (Fe, Co, or Ni) is between 0.10 and 0.25. Therefore, the assumed ratio of 0.131 is justified.} This calculation yields an averaged total magnetic moment $\mu_{t}$ of $\mu_{t}=(0.0325\pm 0.0004)$\,$\mu_{\mathrm{B}}$ per Pt atom in the Pt/Fe bilayer, which is comparable to similar experiments.\cite{Poulopoulos:JAP:2001}

In Pt(10\,nm)/YIG and Pt(7\,nm)/YIG, however, the integrated Pt XMCD signal shows no obvious evidence for induced magnetic moments within the experimental detection limit. In contrast, the integrated XMCD intensity of the Pt(3\,nm)/YIG sample reveals a tiny possible XMCD signal [cf.~black line in Fig.~\ref{fig:Fig2}(c)], which is about ten times smaller than the integrated XMCD signal from the Pt(10\,nm)/Fe bilayer. At present, it is not possible to assess clearly whether this finite integrated intensity is indeed due to a presence of an induced magnetism in the Pt layer, or whether it is an experimental artifact, since the signal magnitude is comparable to the detection limit. Note, however, that if a robust induced polarization would be present in Pt, one would expect integrated XMCD signals from the 7nm and 10nm samples of about one half and one third the amplitude of the 3nm sample signal, which is not consistent with the experimental data in Fig.~\ref{fig:Fig2}(c). Taken together, the $L_{3}$ edge XMCD data from Pt/YIG bilayers compiled in Fig.\,\ref{fig:Fig2} therefore do not allow to rule out the presence of an induced spin polarization in the Pt layer. However, from the integrated XMCD signal in the Pt(3\,nm)/YIG sample, we estimate an upper limit for the induced total magnetic moment of $(0.003\pm 0.001)$\,$\mu_{\mathrm{B}}$ per Pt atom averaged over the layer thickness.\footnote{The number of $5d$ holes taken in the analysis of the Pt/Fe as well as the Pt/YIG bilayer was 1.63.} Note that a field-induced moment due to paramagnetism in Pt was found to be 0.0011\,$\mu_{\mathrm{B}}$ per Pt atom and per tesla.\cite{Bartolome:PRB:2009} Therefore, this contribution is negligible and can not explain our experimental findings. If we consider that only the first layer of the Pt thin film is polarized, then the Pt atoms at the interface would carry an induced moment, which is at least 30 times smaller than that of the Pt atoms at the interface in Pt/Fe bilayers.

This result appears reasonable considering that in NM/FMM heterostructures, (spin polarized) charge carriers can propagate from the ferromagnet into the 'non-magnetic' normal metal and back, such that a finite spin susceptibility and a finite spin polarization builds up close to the interface. In contrast, in NM/FMI heterostructures, charge carriers can not penetrate into the insulating ferromagnetic layer. One might thus naively expect a substantially reduced or even vanishing spin polarization in the 'non-magnetic' metal, in agreement with our experimental observations. However, we would like to emphasize again that the XMCD measurements presented here are sensitive only to the projection of the magnetic moment onto the X-ray beam direction. To draw conclusive statements about induced moments in NM/FMI heterostructures, XMCD measurements sensitive to both the moment along as well as perpendicular to the beam direction are mandatory. Furthermore, more systematic investigations as a function of the NM and the FMI layer thickness would be desirable.

In summary, taking advantage of the element-specific XMCD technique, we have investigated the possible occurrence of induced magnetic moments in Pt films deposited onto the ferromagnetic insulator yttrium iron garnet (YIG). For comparison, we also recorded the XMCD signal of a Pt film deposited onto a ferromagnetic metal (Fe). While a large magnetic dichroism was detected at the Pt $L_{3}$ edge in the Pt(10\,nm)/Fe (NM/FMM) bilayer evidencing the presence of induced magnetic moments in Pt, our data show no evidence of an induced Pt magnetic polarization in Pt/YIG (NM/FMI) bilayers with a Pt thickness of 10\,nm and 7\,nm. 
A small but finite integrated XMCD signal appears to be present in the Pt(3\,nm)/YIG sample. Our data thus show that if a finite moment is induced in the Pt at all, it is at least 30 times smaller than in the corresponding Pt(10\,nm)/Fe reference sample.


\begin{acknowledgments}
We thank A.~Erb for the preparation of the polycrystalline PLD target, T.~Brenninger for technical support, G.~Woltersdorf and Stuart S.~P.~Parkin for stimulating discussions. This work was supported by the European Synchrotron Radiation Facility (ESRF) via project HE-3784, by the Deutsche Forschungsgemeinschaft (DFG) via the priority programme SPP 1538 'Spin-caloric transport' (project GO 944/4-1), and the German Excellence Initiative via the 'Nanosystems Initiative Munich' (NIM).
\end{acknowledgments}

\clearpage

\begin{thebibliography}{33}%
\makeatletter
\providecommand \@ifxundefined [1]{%
 \@ifx{#1\undefined}
}%
\providecommand \@ifnum [1]{%
 \ifnum #1\expandafter \@firstoftwo
 \else \expandafter \@secondoftwo
 \fi
}%
\providecommand \@ifx [1]{%
 \ifx #1\expandafter \@firstoftwo
 \else \expandafter \@secondoftwo
 \fi
}%
\providecommand \natexlab [1]{#1}%
\providecommand \enquote  [1]{``#1''}%
\providecommand \bibnamefont  [1]{#1}%
\providecommand \bibfnamefont [1]{#1}%
\providecommand \citenamefont [1]{#1}%
\providecommand \href@noop [0]{\@secondoftwo}%
\providecommand \href [0]{\begingroup \@sanitize@url \@href}%
\providecommand \@href[1]{\@@startlink{#1}\@@href}%
\providecommand \@@href[1]{\endgroup#1\@@endlink}%
\providecommand \@sanitize@url [0]{\catcode `\\12\catcode `\$12\catcode
  `\&12\catcode `\#12\catcode `\^12\catcode `\_12\catcode `\%12\relax}%
\providecommand \@@startlink[1]{}%
\providecommand \@@endlink[0]{}%
\providecommand \url  [0]{\begingroup\@sanitize@url \@url }%
\providecommand \@url [1]{\endgroup\@href {#1}{\urlprefix }}%
\providecommand \urlprefix  [0]{URL }%
\providecommand \Eprint [0]{\href }%
\providecommand \doibase [0]{http://dx.doi.org/}%
\providecommand \selectlanguage [0]{\@gobble}%
\providecommand \bibinfo  [0]{\@secondoftwo}%
\providecommand \bibfield  [0]{\@secondoftwo}%
\providecommand \translation [1]{[#1]}%
\providecommand \BibitemOpen [0]{}%
\providecommand \bibitemStop [0]{}%
\providecommand \bibitemNoStop [0]{.\EOS\space}%
\providecommand \EOS [0]{\spacefactor3000\relax}%
\providecommand \BibitemShut  [1]{\csname bibitem#1\endcsname}%
\let\auto@bib@innerbib\@empty
\bibitem [{\citenamefont {Hirsch}(1999)}]{Hirsch:PRL:1999}%
  \BibitemOpen
  \bibfield  {author} {\bibinfo {author} {\bibfnamefont {J.~E.}\ \bibnamefont
  {Hirsch}},\ }\href {\doibase 10.1103/PhysRevLett.83.1834} {\bibfield
  {journal} {\bibinfo  {journal} {Phys. Rev. Lett.}\ }\textbf {\bibinfo
  {volume} {83}},\ \bibinfo {pages} {1834} (\bibinfo {year}
  {1999})}\BibitemShut {NoStop}%
\bibitem [{\citenamefont {Stevens}\ \emph {et~al.}(2003)\citenamefont
  {Stevens}, \citenamefont {Smirl}, \citenamefont {Bhat}, \citenamefont
  {Najmaie}, \citenamefont {Sipe},\ and\ \citenamefont {van
  Driel}}]{Stevens:PRL:2003}%
  \BibitemOpen
  \bibfield  {author} {\bibinfo {author} {\bibfnamefont {M.~J.}\ \bibnamefont
  {Stevens}}, \bibinfo {author} {\bibfnamefont {A.~L.}\ \bibnamefont {Smirl}},
  \bibinfo {author} {\bibfnamefont {R.~D.~R.}\ \bibnamefont {Bhat}}, \bibinfo
  {author} {\bibfnamefont {A.}~\bibnamefont {Najmaie}}, \bibinfo {author}
  {\bibfnamefont {J.~E.}\ \bibnamefont {Sipe}}, \ and\ \bibinfo {author}
  {\bibfnamefont {H.~M.}\ \bibnamefont {van Driel}},\ }\href {\doibase
  10.1103/PhysRevLett.90.136603} {\bibfield  {journal} {\bibinfo  {journal}
  {Phys. Rev. Lett.}\ }\textbf {\bibinfo {volume} {90}},\ \bibinfo {pages}
  {136603} (\bibinfo {year} {2003})}\BibitemShut {NoStop}%
\bibitem [{\citenamefont {Kato}\ \emph {et~al.}(2004)\citenamefont {Kato},
  \citenamefont {Myers}, \citenamefont {Gossard},\ and\ \citenamefont
  {Awschalom}}]{Kato:Science:2004}%
  \BibitemOpen
  \bibfield  {author} {\bibinfo {author} {\bibfnamefont {Y.~K.}\ \bibnamefont
  {Kato}}, \bibinfo {author} {\bibfnamefont {R.~C.}\ \bibnamefont {Myers}},
  \bibinfo {author} {\bibfnamefont {A.~C.}\ \bibnamefont {Gossard}}, \ and\
  \bibinfo {author} {\bibfnamefont {D.~D.}\ \bibnamefont {Awschalom}},\ }\href
  {\doibase 10.1126/science.1105514} {\bibfield  {journal} {\bibinfo  {journal}
  {Science}\ }\textbf {\bibinfo {volume} {306}},\ \bibinfo {pages} {1910}
  (\bibinfo {year} {2004})}\BibitemShut {NoStop}%
\bibitem [{\citenamefont {Saitoh}\ \emph {et~al.}(2006)\citenamefont {Saitoh},
  \citenamefont {Ueda}, \citenamefont {Miyajima},\ and\ \citenamefont
  {Tatara}}]{Saitoh:APL:2006}%
  \BibitemOpen
  \bibfield  {author} {\bibinfo {author} {\bibfnamefont {E.}~\bibnamefont
  {Saitoh}}, \bibinfo {author} {\bibfnamefont {M.}~\bibnamefont {Ueda}},
  \bibinfo {author} {\bibfnamefont {H.}~\bibnamefont {Miyajima}}, \ and\
  \bibinfo {author} {\bibfnamefont {G.}~\bibnamefont {Tatara}},\ }\href
  {\doibase 10.1063/1.2199473} {\bibfield  {journal} {\bibinfo  {journal}
  {Appl. Phys. Lett.}\ }\textbf {\bibinfo {volume} {88}},\ \bibinfo {eid}
  {182509} (\bibinfo {year} {2006})}\BibitemShut {NoStop}%
\bibitem [{\citenamefont {Mosendz}\ \emph {et~al.}(2010)\citenamefont
  {Mosendz}, \citenamefont {Pearson}, \citenamefont {Fradin}, \citenamefont
  {Bauer}, \citenamefont {Bader},\ and\ \citenamefont
  {Hoffmann}}]{Mosendz:PRL:2010}%
  \BibitemOpen
  \bibfield  {author} {\bibinfo {author} {\bibfnamefont {O.}~\bibnamefont
  {Mosendz}}, \bibinfo {author} {\bibfnamefont {J.~E.}\ \bibnamefont
  {Pearson}}, \bibinfo {author} {\bibfnamefont {F.~Y.}\ \bibnamefont {Fradin}},
  \bibinfo {author} {\bibfnamefont {G.~E.~W.}\ \bibnamefont {Bauer}}, \bibinfo
  {author} {\bibfnamefont {S.~D.}\ \bibnamefont {Bader}}, \ and\ \bibinfo
  {author} {\bibfnamefont {A.}~\bibnamefont {Hoffmann}},\ }\href {\doibase
  10.1103/PhysRevLett.104.046601} {\bibfield  {journal} {\bibinfo  {journal}
  {Phys. Rev. Lett.}\ }\textbf {\bibinfo {volume} {104}},\ \bibinfo {pages}
  {046601} (\bibinfo {year} {2010})}\BibitemShut {NoStop}%
\bibitem [{\citenamefont {Uchida}\ \emph
  {et~al.}(2010{\natexlab{a}})\citenamefont {Uchida}, \citenamefont {Xiao},
  \citenamefont {Adachi}, \citenamefont {Ohe}, \citenamefont {Takahashi},
  \citenamefont {Ieda}, \citenamefont {Ota}, \citenamefont {Kajiwara},
  \citenamefont {Umezawa}, \citenamefont {Kawai}, \citenamefont {Bauer},
  \citenamefont {Maekawa},\ and\ \citenamefont
  {Saitoh}}]{Uchida:NatMater:2010}%
  \BibitemOpen
  \bibfield  {author} {\bibinfo {author} {\bibfnamefont {K.}~\bibnamefont
  {Uchida}}, \bibinfo {author} {\bibfnamefont {J.}~\bibnamefont {Xiao}},
  \bibinfo {author} {\bibfnamefont {H.}~\bibnamefont {Adachi}}, \bibinfo
  {author} {\bibfnamefont {J.}~\bibnamefont {Ohe}}, \bibinfo {author}
  {\bibfnamefont {S.}~\bibnamefont {Takahashi}}, \bibinfo {author}
  {\bibfnamefont {J.}~\bibnamefont {Ieda}}, \bibinfo {author} {\bibfnamefont
  {T.}~\bibnamefont {Ota}}, \bibinfo {author} {\bibfnamefont {Y.}~\bibnamefont
  {Kajiwara}}, \bibinfo {author} {\bibfnamefont {H.}~\bibnamefont {Umezawa}},
  \bibinfo {author} {\bibfnamefont {H.}~\bibnamefont {Kawai}}, \bibinfo
  {author} {\bibfnamefont {G.~E.~W.}\ \bibnamefont {Bauer}}, \bibinfo {author}
  {\bibfnamefont {S.}~\bibnamefont {Maekawa}}, \ and\ \bibinfo {author}
  {\bibfnamefont {E.}~\bibnamefont {Saitoh}},\ }\href {\doibase
  10.1038/nmat2856} {\bibfield  {journal} {\bibinfo  {journal} {Nat. Mater.}\
  }\textbf {\bibinfo {volume} {9}},\ \bibinfo {pages} {894} (\bibinfo {year}
  {2010}{\natexlab{a}})}\BibitemShut {NoStop}%
\bibitem [{Note1()}]{Note1}%
  \BibitemOpen
  \bibinfo {note} {For the sake of simplicity, we use the term 'ferromagnetic'
  for both ferro- and ferrimagnetic materials throughout this paper. A
  'ferromagnetic insulator' thus is an electrically insulating material which
  exhibits long-range ferro- or ferri-magnetic ordering.}\BibitemShut {Stop}%
\bibitem [{\citenamefont {Tserkovnyak}\ \emph {et~al.}(2002)\citenamefont
  {Tserkovnyak}, \citenamefont {Brataas},\ and\ \citenamefont
  {Bauer}}]{Tserkovnyak:PRL:2002}%
  \BibitemOpen
  \bibfield  {author} {\bibinfo {author} {\bibfnamefont {Y.}~\bibnamefont
  {Tserkovnyak}}, \bibinfo {author} {\bibfnamefont {A.}~\bibnamefont
  {Brataas}}, \ and\ \bibinfo {author} {\bibfnamefont {G.~E.~W.}\ \bibnamefont
  {Bauer}},\ }\href {\doibase 10.1103/PhysRevLett.88.117601} {\bibfield
  {journal} {\bibinfo  {journal} {Phys. Rev. Lett.}\ }\textbf {\bibinfo
  {volume} {88}},\ \bibinfo {pages} {117601} (\bibinfo {year}
  {2002})}\BibitemShut {NoStop}%
\bibitem [{\citenamefont {Czeschka}\ \emph {et~al.}(2011)\citenamefont
  {Czeschka}, \citenamefont {Dreher}, \citenamefont {Brandt}, \citenamefont
  {Weiler}, \citenamefont {Althammer}, \citenamefont {Imort}, \citenamefont
  {Reiss}, \citenamefont {Thomas}, \citenamefont {Schoch}, \citenamefont
  {Limmer}, \citenamefont {Huebl}, \citenamefont {Gross},\ and\ \citenamefont
  {Goennenwein}}]{Czeschka:PRL:2011}%
  \BibitemOpen
  \bibfield  {author} {\bibinfo {author} {\bibfnamefont {F.~D.}\ \bibnamefont
  {Czeschka}}, \bibinfo {author} {\bibfnamefont {L.}~\bibnamefont {Dreher}},
  \bibinfo {author} {\bibfnamefont {M.~S.}\ \bibnamefont {Brandt}}, \bibinfo
  {author} {\bibfnamefont {M.}~\bibnamefont {Weiler}}, \bibinfo {author}
  {\bibfnamefont {M.}~\bibnamefont {Althammer}}, \bibinfo {author}
  {\bibfnamefont {I.-M.}\ \bibnamefont {Imort}}, \bibinfo {author}
  {\bibfnamefont {G.}~\bibnamefont {Reiss}}, \bibinfo {author} {\bibfnamefont
  {A.}~\bibnamefont {Thomas}}, \bibinfo {author} {\bibfnamefont
  {W.}~\bibnamefont {Schoch}}, \bibinfo {author} {\bibfnamefont
  {W.}~\bibnamefont {Limmer}}, \bibinfo {author} {\bibfnamefont
  {H.}~\bibnamefont {Huebl}}, \bibinfo {author} {\bibfnamefont
  {R.}~\bibnamefont {Gross}}, \ and\ \bibinfo {author} {\bibfnamefont
  {S.~T.~B.}\ \bibnamefont {Goennenwein}},\ }\href {\doibase
  10.1103/PhysRevLett.107.046601} {\bibfield  {journal} {\bibinfo  {journal}
  {Phys. Rev. Lett.}\ }\textbf {\bibinfo {volume} {107}},\ \bibinfo {pages}
  {046601} (\bibinfo {year} {2011})}\BibitemShut {NoStop}%
\bibitem [{\citenamefont {Uchida}\ \emph {et~al.}(2008)\citenamefont {Uchida},
  \citenamefont {Takahashi}, \citenamefont {Harii}, \citenamefont {Ieda},
  \citenamefont {Koshibae}, \citenamefont {Ando}, \citenamefont {Maekawa},\
  and\ \citenamefont {Saitoh}}]{Uchida:Nature:2008}%
  \BibitemOpen
  \bibfield  {author} {\bibinfo {author} {\bibfnamefont {K.}~\bibnamefont
  {Uchida}}, \bibinfo {author} {\bibfnamefont {S.}~\bibnamefont {Takahashi}},
  \bibinfo {author} {\bibfnamefont {K.}~\bibnamefont {Harii}}, \bibinfo
  {author} {\bibfnamefont {J.}~\bibnamefont {Ieda}}, \bibinfo {author}
  {\bibfnamefont {W.}~\bibnamefont {Koshibae}}, \bibinfo {author}
  {\bibfnamefont {K.}~\bibnamefont {Ando}}, \bibinfo {author} {\bibfnamefont
  {S.}~\bibnamefont {Maekawa}}, \ and\ \bibinfo {author} {\bibfnamefont
  {E.}~\bibnamefont {Saitoh}},\ }\href {\doibase 10.1038/nature07321}
  {\bibfield  {journal} {\bibinfo  {journal} {Nature}\ }\textbf {\bibinfo
  {volume} {455}},\ \bibinfo {pages} {778} (\bibinfo {year}
  {2008})}\BibitemShut {NoStop}%
\bibitem [{\citenamefont {Jaworski}\ \emph {et~al.}(2010)\citenamefont
  {Jaworski}, \citenamefont {Yang}, \citenamefont {Mack}, \citenamefont
  {Awschalom}, \citenamefont {Heremans},\ and\ \citenamefont
  {Myers}}]{Jaworski:NatMater:2010}%
  \BibitemOpen
  \bibfield  {author} {\bibinfo {author} {\bibfnamefont {C.~M.}\ \bibnamefont
  {Jaworski}}, \bibinfo {author} {\bibfnamefont {J.}~\bibnamefont {Yang}},
  \bibinfo {author} {\bibfnamefont {S.}~\bibnamefont {Mack}}, \bibinfo {author}
  {\bibfnamefont {D.~D.}\ \bibnamefont {Awschalom}}, \bibinfo {author}
  {\bibfnamefont {J.~P.}\ \bibnamefont {Heremans}}, \ and\ \bibinfo {author}
  {\bibfnamefont {R.~C.}\ \bibnamefont {Myers}},\ }\href {\doibase
  10.1038/nmat2860} {\bibfield  {journal} {\bibinfo  {journal} {Nat. Mater.}\
  }\textbf {\bibinfo {volume} {9}},\ \bibinfo {pages} {898} (\bibinfo {year}
  {2010})}\BibitemShut {NoStop}%
\bibitem [{\citenamefont {Weiler}\ \emph {et~al.}(2012)\citenamefont {Weiler},
  \citenamefont {Althammer}, \citenamefont {Czeschka}, \citenamefont {Huebl},
  \citenamefont {Wagner}, \citenamefont {Opel}, \citenamefont {Imort},
  \citenamefont {Reiss}, \citenamefont {Thomas}, \citenamefont {Gross},\ and\
  \citenamefont {Goennenwein}}]{Weiler:PRL:2012}%
  \BibitemOpen
  \bibfield  {author} {\bibinfo {author} {\bibfnamefont {M.}~\bibnamefont
  {Weiler}}, \bibinfo {author} {\bibfnamefont {M.}~\bibnamefont {Althammer}},
  \bibinfo {author} {\bibfnamefont {F.~D.}\ \bibnamefont {Czeschka}}, \bibinfo
  {author} {\bibfnamefont {H.}~\bibnamefont {Huebl}}, \bibinfo {author}
  {\bibfnamefont {M.~S.}\ \bibnamefont {Wagner}}, \bibinfo {author}
  {\bibfnamefont {M.}~\bibnamefont {Opel}}, \bibinfo {author} {\bibfnamefont
  {I.-M.}\ \bibnamefont {Imort}}, \bibinfo {author} {\bibfnamefont
  {G.}~\bibnamefont {Reiss}}, \bibinfo {author} {\bibfnamefont
  {A.}~\bibnamefont {Thomas}}, \bibinfo {author} {\bibfnamefont
  {R.}~\bibnamefont {Gross}}, \ and\ \bibinfo {author} {\bibfnamefont
  {S.~T.~B.}\ \bibnamefont {Goennenwein}},\ }\href {\doibase
  10.1103/PhysRevLett.108.106602} {\bibfield  {journal} {\bibinfo  {journal}
  {Phys. Rev. Lett.}\ }\textbf {\bibinfo {volume} {108}},\ \bibinfo {pages}
  {106602} (\bibinfo {year} {2012})}\BibitemShut {NoStop}%
\bibitem [{\citenamefont {Jaworski}\ \emph {et~al.}(2012)\citenamefont
  {Jaworski}, \citenamefont {Myers}, \citenamefont {Johnston-Halperin},\ and\
  \citenamefont {Heremans}}]{Jaworski:Nature:2012}%
  \BibitemOpen
  \bibfield  {author} {\bibinfo {author} {\bibfnamefont {C.~M.}\ \bibnamefont
  {Jaworski}}, \bibinfo {author} {\bibfnamefont {R.~C.}\ \bibnamefont {Myers}},
  \bibinfo {author} {\bibfnamefont {E.}~\bibnamefont {Johnston-Halperin}}, \
  and\ \bibinfo {author} {\bibfnamefont {J.~P.}\ \bibnamefont {Heremans}},\
  }\href {\doibase 10.1038/nature11221} {\bibfield  {journal} {\bibinfo
  {journal} {Nature}\ }\textbf {\bibinfo {volume} {487}},\ \bibinfo {pages}
  {210} (\bibinfo {year} {2012})}\BibitemShut {NoStop}%
\bibitem [{\citenamefont {Uchida}\ \emph
  {et~al.}(2010{\natexlab{b}})\citenamefont {Uchida}, \citenamefont {Adachi},
  \citenamefont {Ota}, \citenamefont {Nakayama}, \citenamefont {Maekawa},\ and\
  \citenamefont {Saitoh}}]{Uchida:APL:2010}%
  \BibitemOpen
  \bibfield  {author} {\bibinfo {author} {\bibfnamefont {K.}~\bibnamefont
  {Uchida}}, \bibinfo {author} {\bibfnamefont {H.}~\bibnamefont {Adachi}},
  \bibinfo {author} {\bibfnamefont {T.}~\bibnamefont {Ota}}, \bibinfo {author}
  {\bibfnamefont {H.}~\bibnamefont {Nakayama}}, \bibinfo {author}
  {\bibfnamefont {S.}~\bibnamefont {Maekawa}}, \ and\ \bibinfo {author}
  {\bibfnamefont {E.}~\bibnamefont {Saitoh}},\ }\href {\doibase
  10.1063/1.3507386} {\bibfield  {journal} {\bibinfo  {journal} {Appl. Phys.
  Lett.}\ }\textbf {\bibinfo {volume} {97}},\ \bibinfo {eid} {172505} (\bibinfo
  {year} {2010}{\natexlab{b}})}\BibitemShut {NoStop}%
\bibitem [{\citenamefont {Huang}\ \emph {et~al.}(2012)\citenamefont {Huang},
  \citenamefont {Fan}, \citenamefont {Qu}, \citenamefont {Chen}, \citenamefont
  {Wang}, \citenamefont {Wu}, \citenamefont {Chen}, \citenamefont {Xiao},\ and\
  \citenamefont {Chien}}]{Huang:PRL:2012}%
  \BibitemOpen
  \bibfield  {author} {\bibinfo {author} {\bibfnamefont {S.~Y.}\ \bibnamefont
  {Huang}}, \bibinfo {author} {\bibfnamefont {X.}~\bibnamefont {Fan}}, \bibinfo
  {author} {\bibfnamefont {D.}~\bibnamefont {Qu}}, \bibinfo {author}
  {\bibfnamefont {Y.~P.}\ \bibnamefont {Chen}}, \bibinfo {author}
  {\bibfnamefont {W.~G.}\ \bibnamefont {Wang}}, \bibinfo {author}
  {\bibfnamefont {J.}~\bibnamefont {Wu}}, \bibinfo {author} {\bibfnamefont
  {T.~Y.}\ \bibnamefont {Chen}}, \bibinfo {author} {\bibfnamefont {J.~Q.}\
  \bibnamefont {Xiao}}, \ and\ \bibinfo {author} {\bibfnamefont {C.~L.}\
  \bibnamefont {Chien}},\ }\href {\doibase 10.1103/PhysRevLett.109.107204}
  {\bibfield  {journal} {\bibinfo  {journal} {Phys. Rev. Lett.}\ }\textbf
  {\bibinfo {volume} {109}},\ \bibinfo {pages} {107204} (\bibinfo {year}
  {2012})}\BibitemShut {NoStop}%
\bibitem [{\citenamefont {Wilhelm}\ \emph {et~al.}(2000)\citenamefont
  {Wilhelm}, \citenamefont {Poulopoulos}, \citenamefont {Ceballos},
  \citenamefont {Wende}, \citenamefont {Baberschke}, \citenamefont
  {Srivastava}, \citenamefont {Benea}, \citenamefont {Ebert}, \citenamefont
  {Angelakeris}, \citenamefont {Flevaris}, \citenamefont {Niarchos},
  \citenamefont {Rogalev},\ and\ \citenamefont {Brookes}}]{Wilhelm:PRL:2000}%
  \BibitemOpen
  \bibfield  {author} {\bibinfo {author} {\bibfnamefont {F.}~\bibnamefont
  {Wilhelm}}, \bibinfo {author} {\bibfnamefont {P.}~\bibnamefont
  {Poulopoulos}}, \bibinfo {author} {\bibfnamefont {G.}~\bibnamefont
  {Ceballos}}, \bibinfo {author} {\bibfnamefont {H.}~\bibnamefont {Wende}},
  \bibinfo {author} {\bibfnamefont {K.}~\bibnamefont {Baberschke}}, \bibinfo
  {author} {\bibfnamefont {P.}~\bibnamefont {Srivastava}}, \bibinfo {author}
  {\bibfnamefont {D.}~\bibnamefont {Benea}}, \bibinfo {author} {\bibfnamefont
  {H.}~\bibnamefont {Ebert}}, \bibinfo {author} {\bibfnamefont
  {M.}~\bibnamefont {Angelakeris}}, \bibinfo {author} {\bibfnamefont {N.~K.}\
  \bibnamefont {Flevaris}}, \bibinfo {author} {\bibfnamefont {D.}~\bibnamefont
  {Niarchos}}, \bibinfo {author} {\bibfnamefont {A.}~\bibnamefont {Rogalev}}, \
  and\ \bibinfo {author} {\bibfnamefont {N.~B.}\ \bibnamefont {Brookes}},\
  }\href {\doibase 10.1103/PhysRevLett.85.413} {\bibfield  {journal} {\bibinfo
  {journal} {Phys. Rev. Lett.}\ }\textbf {\bibinfo {volume} {85}},\ \bibinfo
  {pages} {413} (\bibinfo {year} {2000})}\BibitemShut {NoStop}%
\bibitem [{\citenamefont {Wilhelm}\ \emph {et~al.}(2001)\citenamefont
  {Wilhelm}, \citenamefont {Poulopoulos}, \citenamefont {Wende}, \citenamefont
  {Scherz}, \citenamefont {Baberschke}, \citenamefont {Angelakeris},
  \citenamefont {Flevaris},\ and\ \citenamefont {Rogalev}}]{Wilhelm:PRL:2001}%
  \BibitemOpen
  \bibfield  {author} {\bibinfo {author} {\bibfnamefont {F.}~\bibnamefont
  {Wilhelm}}, \bibinfo {author} {\bibfnamefont {P.}~\bibnamefont
  {Poulopoulos}}, \bibinfo {author} {\bibfnamefont {H.}~\bibnamefont {Wende}},
  \bibinfo {author} {\bibfnamefont {A.}~\bibnamefont {Scherz}}, \bibinfo
  {author} {\bibfnamefont {K.}~\bibnamefont {Baberschke}}, \bibinfo {author}
  {\bibfnamefont {M.}~\bibnamefont {Angelakeris}}, \bibinfo {author}
  {\bibfnamefont {N.~K.}\ \bibnamefont {Flevaris}}, \ and\ \bibinfo {author}
  {\bibfnamefont {A.}~\bibnamefont {Rogalev}},\ }\href {\doibase
  10.1103/PhysRevLett.87.207202} {\bibfield  {journal} {\bibinfo  {journal}
  {Phys. Rev. Lett.}\ }\textbf {\bibinfo {volume} {87}},\ \bibinfo {pages}
  {207202} (\bibinfo {year} {2001})}\BibitemShut {NoStop}%
\bibitem [{\citenamefont {Wilhelm}\ \emph {et~al.}(2004)\citenamefont
  {Wilhelm}, \citenamefont {Angelakeris}, \citenamefont {Jaouen}, \citenamefont
  {Poulopoulos}, \citenamefont {Papaioannou}, \citenamefont {Mueller},
  \citenamefont {Fumagalli}, \citenamefont {Rogalev},\ and\ \citenamefont
  {Flevaris}}]{Wilhelm:PRB:2004}%
  \BibitemOpen
  \bibfield  {author} {\bibinfo {author} {\bibfnamefont {F.}~\bibnamefont
  {Wilhelm}}, \bibinfo {author} {\bibfnamefont {M.}~\bibnamefont
  {Angelakeris}}, \bibinfo {author} {\bibfnamefont {N.}~\bibnamefont {Jaouen}},
  \bibinfo {author} {\bibfnamefont {P.}~\bibnamefont {Poulopoulos}}, \bibinfo
  {author} {\bibfnamefont {E.~T.}\ \bibnamefont {Papaioannou}}, \bibinfo
  {author} {\bibfnamefont {C.}~\bibnamefont {Mueller}}, \bibinfo {author}
  {\bibfnamefont {P.}~\bibnamefont {Fumagalli}}, \bibinfo {author}
  {\bibfnamefont {A.}~\bibnamefont {Rogalev}}, \ and\ \bibinfo {author}
  {\bibfnamefont {N.~K.}\ \bibnamefont {Flevaris}},\ }\href {\doibase
  10.1103/PhysRevB.69.220404} {\bibfield  {journal} {\bibinfo  {journal} {Phys.
  Rev. B}\ }\textbf {\bibinfo {volume} {69}},\ \bibinfo {pages} {220404}
  (\bibinfo {year} {2004})}\BibitemShut {NoStop}%
\bibitem [{\citenamefont {Wende}(2004)}]{Wende:RepProgPhys:2004}%
  \BibitemOpen
  \bibfield  {author} {\bibinfo {author} {\bibfnamefont {H.}~\bibnamefont
  {Wende}},\ }\href {\doibase 10.1088/0034-4885/67/12/R01} {\bibfield
  {journal} {\bibinfo  {journal} {Rep. Prog. Phys.}\ }\textbf {\bibinfo
  {volume} {67}},\ \bibinfo {pages} {2105} (\bibinfo {year}
  {2004})}\BibitemShut {NoStop}%
\bibitem [{\citenamefont {Parra}\ and\ \citenamefont
  {Medina}(1980)}]{Parra:PRB:1980}%
  \BibitemOpen
  \bibfield  {author} {\bibinfo {author} {\bibfnamefont {R.~E.}\ \bibnamefont
  {Parra}}\ and\ \bibinfo {author} {\bibfnamefont {R.}~\bibnamefont {Medina}},\
  }\href {\doibase 10.1103/PhysRevB.22.5460} {\bibfield  {journal} {\bibinfo
  {journal} {Phys. Rev. B}\ }\textbf {\bibinfo {volume} {22}},\ \bibinfo
  {pages} {5460} (\bibinfo {year} {1980})}\BibitemShut {NoStop}%
\bibitem [{\citenamefont {Coey}(2010)}]{Coey:book}%
  \BibitemOpen
  \bibfield  {author} {\bibinfo {author} {\bibfnamefont {J.~M.~D.}\
  \bibnamefont {Coey}},\ }\href@noop {} {\bibinfo {title} {Magnetism and
  Magnetic Materials}}\ (\bibinfo  {publisher} {Cambridge University Press},\
  \bibinfo {year} {2010})\BibitemShut {NoStop}%
\bibitem [{\citenamefont {Popova}\ \emph {et~al.}(2003)\citenamefont {Popova},
  \citenamefont {Keller}, \citenamefont {Jomard}, \citenamefont {Thomas},
  \citenamefont {Brianso}, \citenamefont {Gendron}, \citenamefont {Guyot},\
  and\ \citenamefont {Tessier}}]{Popova:EPJB:2003}%
  \BibitemOpen
  \bibfield  {author} {\bibinfo {author} {\bibfnamefont {E.}~\bibnamefont
  {Popova}}, \bibinfo {author} {\bibfnamefont {N.}~\bibnamefont {Keller}},
  \bibinfo {author} {\bibfnamefont {F.}~\bibnamefont {Jomard}}, \bibinfo
  {author} {\bibfnamefont {L.}~\bibnamefont {Thomas}}, \bibinfo {author}
  {\bibfnamefont {M.-C.}\ \bibnamefont {Brianso}}, \bibinfo {author}
  {\bibfnamefont {F.}~\bibnamefont {Gendron}}, \bibinfo {author} {\bibfnamefont
  {M.}~\bibnamefont {Guyot}}, \ and\ \bibinfo {author} {\bibfnamefont
  {M.}~\bibnamefont {Tessier}},\ }\href {\doibase 10.1140/epjb/e2003-00010-2}
  {\bibfield  {journal} {\bibinfo  {journal} {Eur. Phys. J. B}\ }\textbf
  {\bibinfo {volume} {31}},\ \bibinfo {pages} {69} (\bibinfo {year}
  {2003})}\BibitemShut {NoStop}%
\bibitem [{\citenamefont {Rogalev}\ \emph {et~al.}(2009)\citenamefont
  {Rogalev}, \citenamefont {Goulon}, \citenamefont {Wilhelm}, \citenamefont
  {Brouder}, \citenamefont {Yaresko}, \citenamefont {Ben~Youssef},\ and\
  \citenamefont {Indenbom}}]{Rogalev:JMMM:2009}%
  \BibitemOpen
  \bibfield  {author} {\bibinfo {author} {\bibfnamefont {A.}~\bibnamefont
  {Rogalev}}, \bibinfo {author} {\bibfnamefont {J.}~\bibnamefont {Goulon}},
  \bibinfo {author} {\bibfnamefont {F.}~\bibnamefont {Wilhelm}}, \bibinfo
  {author} {\bibfnamefont {C.}~\bibnamefont {Brouder}}, \bibinfo {author}
  {\bibfnamefont {A.}~\bibnamefont {Yaresko}}, \bibinfo {author} {\bibfnamefont
  {J.}~\bibnamefont {Ben~Youssef}}, \ and\ \bibinfo {author} {\bibfnamefont
  {M.}~\bibnamefont {Indenbom}},\ }\href {\doibase 10.1016/j.jmmm.2009.07.078}
  {\bibfield  {journal} {\bibinfo  {journal} {J. Magn. Magn. Mater.}\ }\textbf
  {\bibinfo {volume} {321}},\ \bibinfo {pages} {3945} (\bibinfo {year}
  {2009})}\BibitemShut {NoStop}%
\bibitem [{\citenamefont {Kawamura}\ \emph {et~al.}(1997)\citenamefont
  {Kawamura}, \citenamefont {Maruyama}, \citenamefont {Kobayashi},
  \citenamefont {Inoue},\ and\ \citenamefont
  {Yamazaki}}]{Kawamura:JPhysIV:1997}%
  \BibitemOpen
  \bibfield  {author} {\bibinfo {author} {\bibfnamefont {N.}~\bibnamefont
  {Kawamura}}, \bibinfo {author} {\bibfnamefont {H.}~\bibnamefont {Maruyama}},
  \bibinfo {author} {\bibfnamefont {K.}~\bibnamefont {Kobayashi}}, \bibinfo
  {author} {\bibfnamefont {I.}~\bibnamefont {Inoue}}, \ and\ \bibinfo {author}
  {\bibfnamefont {H.}~\bibnamefont {Yamazaki}},\ }\href {\doibase
  10.1051/jp4:19971105} {\bibfield  {journal} {\bibinfo  {journal} {J. Phys.
  IV}\ }\textbf {\bibinfo {volume} {07}},\ \bibinfo {pages} {C1} (\bibinfo
  {year} {1997})}\BibitemShut {NoStop}%
\bibitem [{\citenamefont {H.~Maruyama}(2004)}]{Maruyama:ElecSpec:2004}%
  \BibitemOpen
  \bibfield  {author} {\bibinfo {author} {\bibfnamefont {N.~K.}\ \bibnamefont
  {H.~Maruyama}},\ }\href {\doibase 10.1016/j.elspec.2004.02.140} {\bibfield
  {journal} {\bibinfo  {journal} {J. Electron Spectrosc. Relat. Phenom.}\
  }\textbf {\bibinfo {volume} {136}},\ \bibinfo {pages} {135} (\bibinfo {year}
  {2004})}\BibitemShut {NoStop}%
\bibitem [{\citenamefont {Mansour}\ \emph {et~al.}(1984)\citenamefont
  {Mansour}, \citenamefont {Sayers}, \citenamefont {Cook}, \citenamefont
  {Short}, \citenamefont {Shannon},\ and\ \citenamefont
  {Katzer}}]{Mansour:JPhysChem:1984}%
  \BibitemOpen
  \bibfield  {author} {\bibinfo {author} {\bibfnamefont {A.~N.}\ \bibnamefont
  {Mansour}}, \bibinfo {author} {\bibfnamefont {D.~E.}\ \bibnamefont {Sayers}},
  \bibinfo {author} {\bibfnamefont {J.~W.}\ \bibnamefont {Cook}}, \bibinfo
  {author} {\bibfnamefont {D.~R.}\ \bibnamefont {Short}}, \bibinfo {author}
  {\bibfnamefont {R.~D.}\ \bibnamefont {Shannon}}, \ and\ \bibinfo {author}
  {\bibfnamefont {J.~R.}\ \bibnamefont {Katzer}},\ }\href {\doibase
  10.1021/j150653a022} {\bibfield  {journal} {\bibinfo  {journal} {J. Phys.
  Chem.}\ }\textbf {\bibinfo {volume} {88}},\ \bibinfo {pages} {1778} (\bibinfo
  {year} {1984})}\BibitemShut {NoStop}%
\bibitem [{\citenamefont {Friebel}\ \emph {et~al.}(2011)\citenamefont
  {Friebel}, \citenamefont {Miller}, \citenamefont {{O'Grady}}, \citenamefont
  {Anniyev}, \citenamefont {Bargar}, \citenamefont {Bergmann}, \citenamefont
  {Ogasawara}, \citenamefont {Wikfeldt}, \citenamefont {Pettersson},\ and\
  \citenamefont {Nilsson}}]{Friebel:PhysChemChemPhys:2011}%
  \BibitemOpen
  \bibfield  {author} {\bibinfo {author} {\bibfnamefont {D.}~\bibnamefont
  {Friebel}}, \bibinfo {author} {\bibfnamefont {D.~J.}\ \bibnamefont {Miller}},
  \bibinfo {author} {\bibfnamefont {C.~P.}\ \bibnamefont {{O'Grady}}}, \bibinfo
  {author} {\bibfnamefont {T.}~\bibnamefont {Anniyev}}, \bibinfo {author}
  {\bibfnamefont {J.}~\bibnamefont {Bargar}}, \bibinfo {author} {\bibfnamefont
  {U.}~\bibnamefont {Bergmann}}, \bibinfo {author} {\bibfnamefont
  {H.}~\bibnamefont {Ogasawara}}, \bibinfo {author} {\bibfnamefont {K.~T.}\
  \bibnamefont {Wikfeldt}}, \bibinfo {author} {\bibfnamefont {L.~G.~M.}\
  \bibnamefont {Pettersson}}, \ and\ \bibinfo {author} {\bibfnamefont
  {A.}~\bibnamefont {Nilsson}},\ }\href {\doibase 10.1039/c0cp01434f}
  {\bibfield  {journal} {\bibinfo  {journal} {Phys. Chem. Chem. Phys.}\
  }\textbf {\bibinfo {volume} {13}},\ \bibinfo {pages} {262} (\bibinfo {year}
  {2011})}\BibitemShut {NoStop}%
\bibitem [{\citenamefont {Miller}\ \emph {et~al.}(2011)\citenamefont {Miller},
  \citenamefont {\"Oberg}, \citenamefont {Kaya}, \citenamefont
  {Sanchez~Casalongue}, \citenamefont {Friebel}, \citenamefont {Anniyev},
  \citenamefont {Ogasawara}, \citenamefont {Bluhm}, \citenamefont
  {Pettersson},\ and\ \citenamefont {Nilsson}}]{Miller:PRL:2011}%
  \BibitemOpen
  \bibfield  {author} {\bibinfo {author} {\bibfnamefont {D.~J.}\ \bibnamefont
  {Miller}}, \bibinfo {author} {\bibfnamefont {H.}~\bibnamefont {\"Oberg}},
  \bibinfo {author} {\bibfnamefont {S.}~\bibnamefont {Kaya}}, \bibinfo {author}
  {\bibfnamefont {H.}~\bibnamefont {Sanchez~Casalongue}}, \bibinfo {author}
  {\bibfnamefont {D.}~\bibnamefont {Friebel}}, \bibinfo {author} {\bibfnamefont
  {T.}~\bibnamefont {Anniyev}}, \bibinfo {author} {\bibfnamefont
  {H.}~\bibnamefont {Ogasawara}}, \bibinfo {author} {\bibfnamefont
  {H.}~\bibnamefont {Bluhm}}, \bibinfo {author} {\bibfnamefont {L.~G.~M.}\
  \bibnamefont {Pettersson}}, \ and\ \bibinfo {author} {\bibfnamefont
  {A.}~\bibnamefont {Nilsson}},\ }\href {\doibase
  10.1103/PhysRevLett.107.195502} {\bibfield  {journal} {\bibinfo  {journal}
  {Phys. Rev. Lett.}\ }\textbf {\bibinfo {volume} {107}},\ \bibinfo {pages}
  {195502} (\bibinfo {year} {2011})}\BibitemShut {NoStop}%
\bibitem [{\citenamefont {Antoniak}\ \emph {et~al.}(2006)\citenamefont
  {Antoniak}, \citenamefont {Lindner}, \citenamefont {Spasova}, \citenamefont
  {Sudfeld}, \citenamefont {Acet}, \citenamefont {Farle}, \citenamefont
  {Fauth}, \citenamefont {Wiedwald}, \citenamefont {Boyen}, \citenamefont
  {Ziemann}, \citenamefont {Wilhelm}, \citenamefont {Rogalev},\ and\
  \citenamefont {Sun}}]{Antoniak:PRL:2006}%
  \BibitemOpen
  \bibfield  {author} {\bibinfo {author} {\bibfnamefont {C.}~\bibnamefont
  {Antoniak}}, \bibinfo {author} {\bibfnamefont {J.}~\bibnamefont {Lindner}},
  \bibinfo {author} {\bibfnamefont {M.}~\bibnamefont {Spasova}}, \bibinfo
  {author} {\bibfnamefont {D.}~\bibnamefont {Sudfeld}}, \bibinfo {author}
  {\bibfnamefont {M.}~\bibnamefont {Acet}}, \bibinfo {author} {\bibfnamefont
  {M.}~\bibnamefont {Farle}}, \bibinfo {author} {\bibfnamefont
  {K.}~\bibnamefont {Fauth}}, \bibinfo {author} {\bibfnamefont
  {U.}~\bibnamefont {Wiedwald}}, \bibinfo {author} {\bibfnamefont {H.-G.}\
  \bibnamefont {Boyen}}, \bibinfo {author} {\bibfnamefont {P.}~\bibnamefont
  {Ziemann}}, \bibinfo {author} {\bibfnamefont {F.}~\bibnamefont {Wilhelm}},
  \bibinfo {author} {\bibfnamefont {A.}~\bibnamefont {Rogalev}}, \ and\
  \bibinfo {author} {\bibfnamefont {S.}~\bibnamefont {Sun}},\ }\href {\doibase
  10.1103/PhysRevLett.97.117201} {\bibfield  {journal} {\bibinfo  {journal}
  {Phys. Rev. Lett.}\ }\textbf {\bibinfo {volume} {97}},\ \bibinfo {pages}
  {117201} (\bibinfo {year} {2006})}\BibitemShut {NoStop}%
\bibitem [{\citenamefont {Poulopoulos}\ \emph {et~al.}(2001)\citenamefont
  {Poulopoulos}, \citenamefont {Wilhelm}, \citenamefont {Wende}, \citenamefont
  {Ceballos}, \citenamefont {Baberschke}, \citenamefont {Benea}, \citenamefont
  {Ebert}, \citenamefont {Angelakeris}, \citenamefont {Flevaris}, \citenamefont
  {Rogalev},\ and\ \citenamefont {Brookes}}]{Poulopoulos:JAP:2001}%
  \BibitemOpen
  \bibfield  {author} {\bibinfo {author} {\bibfnamefont {P.}~\bibnamefont
  {Poulopoulos}}, \bibinfo {author} {\bibfnamefont {F.}~\bibnamefont
  {Wilhelm}}, \bibinfo {author} {\bibfnamefont {H.}~\bibnamefont {Wende}},
  \bibinfo {author} {\bibfnamefont {G.}~\bibnamefont {Ceballos}}, \bibinfo
  {author} {\bibfnamefont {K.}~\bibnamefont {Baberschke}}, \bibinfo {author}
  {\bibfnamefont {D.}~\bibnamefont {Benea}}, \bibinfo {author} {\bibfnamefont
  {H.}~\bibnamefont {Ebert}}, \bibinfo {author} {\bibfnamefont
  {M.}~\bibnamefont {Angelakeris}}, \bibinfo {author} {\bibfnamefont {N.~K.}\
  \bibnamefont {Flevaris}}, \bibinfo {author} {\bibfnamefont {A.}~\bibnamefont
  {Rogalev}}, \ and\ \bibinfo {author} {\bibfnamefont {N.~B.}\ \bibnamefont
  {Brookes}},\ }\href {\doibase 10.1063/1.1345862} {\bibfield  {journal}
  {\bibinfo  {journal} {J. Appl. Phys.}\ }\textbf {\bibinfo {volume} {89}},\
  \bibinfo {pages} {3874} (\bibinfo {year} {2001})}\BibitemShut {NoStop}%
\bibitem [{Note2()}]{Note2}%
  \BibitemOpen
  \bibinfo {note} {The ratio $\mu _{l}/\mu _{s}$ for induced moments in Pt in
  direct contact with a ferromagnetic late $3d$ transition metal (Fe, Co, or
  Ni) is between 0.10 and 0.25. Therefore, the assumed ratio of 0.131 is
  justified.}\BibitemShut {Stop}%
\bibitem [{Note3()}]{Note3}%
  \BibitemOpen
  \bibinfo {note} {The number of $5d$ holes taken in the analysis of the Pt/Fe
  as well as the Pt/YIG bilayer was 1.63.}\BibitemShut {Stop}%
\bibitem [{\citenamefont {Bartolom\'e}\ \emph {et~al.}(2009)\citenamefont
  {Bartolom\'e}, \citenamefont {Bartolom\'e}, \citenamefont {Garc\'ia},
  \citenamefont {Roduner}, \citenamefont {Akdogan}, \citenamefont {Wilhelm},\
  and\ \citenamefont {Rogalev}}]{Bartolome:PRB:2009}%
  \BibitemOpen
  \bibfield  {author} {\bibinfo {author} {\bibfnamefont {J.}~\bibnamefont
  {Bartolom\'e}}, \bibinfo {author} {\bibfnamefont {F.}~\bibnamefont
  {Bartolom\'e}}, \bibinfo {author} {\bibfnamefont {L.~M.}\ \bibnamefont
  {Garc\'ia}}, \bibinfo {author} {\bibfnamefont {E.}~\bibnamefont {Roduner}},
  \bibinfo {author} {\bibfnamefont {Y.}~\bibnamefont {Akdogan}}, \bibinfo
  {author} {\bibfnamefont {F.}~\bibnamefont {Wilhelm}}, \ and\ \bibinfo
  {author} {\bibfnamefont {A.}~\bibnamefont {Rogalev}},\ }\href {\doibase
  10.1103/PhysRevB.80.014404} {\bibfield  {journal} {\bibinfo  {journal} {Phys.
  Rev. B}\ }\textbf {\bibinfo {volume} {80}},\ \bibinfo {pages} {014404}
  (\bibinfo {year} {2009})}\BibitemShut {NoStop}%
\end{thebibliography}
\end{document}